\documentclass[aps,prd,superscriptaddress,floatfix,nofootinbib,nolongbibliography,10pt,twocolumn,preprintnumbers]{revtex4-2}

\usepackage{amsfonts}
\usepackage{amsmath}
\usepackage{amssymb}
\usepackage{bm}
\usepackage{physics}
\usepackage{dsfont}

\usepackage{graphicx}

\usepackage{xcolor}
\usepackage[colorlinks=true,citecolor=blue,linkcolor=blue,urlcolor=blue]{hyperref}

\begin{document}
    
\title{Mapping data sensitivities in global QCD analysis with \\ linear response and influence functions}

\author{R.~M.~Whitehill}
\email{rwhit058@odu.edu}
\affiliation{Department of Physics, Old Dominion University, Norfolk, Virginia 23529, USA}
\affiliation{Jefferson Lab, Newport News, Virginia 23606, USA}

\preprint{JLAB-THY-26-4945}

\begin{abstract}
    Global QCD analyses provide the primary framework for extracting hadron structure from experimental data, yet the mechanisms by which data constrain non-perturbative functions are difficult to quantify and interpret due to the complexity of these fits.
    Here we adapt linear response and influence functions, which are gradient-based measures of the local sensitivity of fitted quantities to perturbations of the data, for use within Hessian and Bayesian global QCD analyses.
    Through a controlled proof of principle, we demonstrate that these quantities reveal how data locally shape the central values, uncertainties, and correlations of quantum correlation functions, providing an interpretable framework for diagnosing information flow in inverse problems in QCD.
\end{abstract}

\maketitle

\section{Introduction}
\label{sec:introduction}

Global QCD analysis provides a robust framework for extracting the internal structure of hadrons from experimental measurements.
In particular, it exploits the universality of quantum correlation functions (QCFs) to solve inverse problems defined by factorization theorems, which relate experimental observables to convolutions of perturbatively calculable functions and the non-perturbative QCFs~\cite{Collins:2011zzd,Collins:1989gx}.
QCFs such as collinear parton distribution functions (PDFs)~\cite{Cocuzza:2026vey,Cocuzza:2025qvf,Hou:2019efy,NNPDF:2021njg,Bailey:2020ooq,Borsa:2024mss,Cerutti:2025yji}, fragmentation functions (FFs)~\cite{Anderson:2024evk,deFlorian:2007aj,Khalek:2021gxf}, transverse momentum dependent distributions (TMDs)~\cite{Barry:2025glq,Bacchetta:2025ara,Moos:2025sal,Aslan:2024nqg}, and more recently generalized parton distributions (GPDs)~\cite{Guo:2025muf,Dotson:2025omi,Panjsheeri:2025vpa} have been determined through increasingly precise and expansive sets of experimental data and sustained theoretical developments.

These analyses integrate multiple interacting theoretical, phenomenological, statistical, and computational ingredients.
For example, the perturbative order of hard scattering functions and evolution kernels~\cite{Bailey:2020ooq,Ball:2011uy,NNPDF:2024dpb,Ball:2012cx}, the treatment of higher-twist or power-suppressed effects~\cite{Cerutti:2025yji,Yang:1999xg,Owens:2012bv,Harland-Lang:2025wvm,Accardi:2016qay,Accardi:2009br,Accardi:2011fa}, the parameterization and modeling of non-perturbative functions~\cite{Bacchetta:2025ara,Kotz:2025une,Ablat:2025gbp}, the imposition of theoretical and phenomenological constraints~\cite{Collins:2021vke,Zhou:2022wzm,Candido:2020yat,DAlesio:2020vtw,Gamberg:2022kdb}, and the statistical and numerical methods employed~\cite{Pumplin:2001ct,Ball:2008by,Risse:2025qlo,NNPDF:2014otw,NNPDF:2017mvq,Salam:2008qg,Watt:2012tq,Carli:2010rw,Stratmann:2001pb} all non-trivially modify the behavior of extracted QCFs and the geometry of their posterior distributions.
While experimental data ultimately constrain the extracted QCFs, the interplay of these layers obscures how different measurements influence specific features of the QCFs.
This becomes especially problematic in modern analyses which couple and simultaneously extract distinct QCFs~\cite{Barry:2025glq}.

Clarifying the underlying connection between the results of a global analysis and data is essential for interpreting fits and understanding how data pulls on a phenomenological model.
Several approaches have been developed to assess data sensitivity and impact, including Lagrange multiplier scans, correlation-based measures, $L_2$ sensitivities, and principal component analyses~\cite{Wang:2018heo,Hobbs:2019gob,Jing:2023isu}.
While these methods have proven valuable in practical applications, their interpretation depends on the underlying formulation and specific representation of the inference problem~\cite{Kriesten:2024are,Bonnet-Guerrini:2026thq}.
Depending on the construction, some characterize sensitivity through variations of the phenomenological model in the forward relation between QCFs and experimental observables.
Some have direct interpretations only in linearized parameter space or provide integrated, high-dimensional summaries that can be difficult to interpret in terms of the fundamental theoretical variables or experimental kinematics.
Here we instead organize sensitivity around the inverse mapping from experimental data to extracted QCFs, asking: \textit{How do perturbations of a data point or dataset propagate to QCFs?}

We address this question by adapting linear response and influence functions for use in Hessian and Bayesian global QCD analyses.
These concepts have roots in robust estimation, classical sensitivity analysis, and linear response theory and have found applications in areas such as machine learning~\cite{Hampel:1974,Cook:1977,Ruelle_2009,giordano2015linear,giordano2019swiss,koh2017understanding}.
Linear response and influence functions, respectively, quantify the local sensitivity of fitted quantities to perturbations of experimental inputs or the statistical weight of data points or datasets.
In this work, we demonstrate and validate this framework using a controlled extraction of unpolarized collinear PDFs from pseudodata generated within a parton-model description of the $F_2$ deep-inelastic scattering (DIS) structure function.
We show that response and influence functions provide a unified framework to probe the data-level sensitivities of QCF central values, uncertainties, and correlations extracted through global QCD analysis.

\section{Overview of global analysis}
\label{sec:overview-of-global-analysis}

The basic setup for a fit, or more generally an inverse problem, is as follows.
Suppose we have an observable $\mathcal{O}$, dependent on a collection of independent variables $\vb*{x}$.
Furthermore, we assume this observable can be described within a theoretical model through a function $T$ which depends on both the independent variables $\vb*{x}$ and model parameters $\vb*{a}$, that is, $T(\vb*{x};\vb*{a})$.
If the parameters $\vb*{a}$ cannot be computed from first principles, we infer them by comparing model predictions to experimental measurements of the observable $\mathcal{O}$.

An experiment yields a dataset $\mathcal{D} \!=\! \big( \{ (\vb*{x}^{(i)},D_{i} ) \} , V \big)$ comprised of measurements $D_j$ of $\mathcal{O}$ at $\vb*{x}^{(i)}$ and a data covariance matrix $V$ enumerating the statistical and systematic error budget of the experiment.
We implicitly assume that the experimental measurements are drawn from a multivariate normal distribution, that is, $\vb*{D} \!\sim\! \mathcal{N}(\expval{ \vb*{\mathcal{O}}(\vb*{x}) },V)$, where $\expval{ \vb*{\mathcal{O}}(\vb*{x}) }$ denotes the collection of true (albeit unknown) values of the observable at the points $\vb*{x}^{(i)}$.
Under the assumption that the theoretical model provides an adequate description of the underlying physics, the true values $\expval{ \mathcal{O}(\vb*{x}) } \!\approx\! T(\vb*{x};\vb*{a})$ for some set of parameters $\vb*{a}$, and the probability of observing the data for a given set of model parameters $\vb*{a}$ is given by the likelihood $\mathcal{L}(\mathcal{D},\vb*{a}) \!\sim\! \exp\big( -\!\chi^2 / 2 \big)$, where $\chi^2 \!=\! \vb*{\Delta}^{T} V^{-1} \vb*{\Delta}$ and $\Delta_{i} \!=\! D_{i} - T(\vb*{x}^{(i)};\vb*{a})$.
Generally, our goal is to identify the parameters which maximize the likelihood or, equivalently, minimize the $\chi^2$, together with a description of the local geometry of the likelihood around this optimum, which encodes uncertainties and correlations among the parameters and quantities derived from them.

\section{Response and influence functions in global QCD analysis}

In this section, we review the definitions of linear response and influence functions within the context of global analyses, as well as some issues relevant to their interpretations and practical application and computation.

\subsection{Response functions}
\label{ssec:response-functions}

The behavior of QCFs extracted from a global analysis depends on the properties of the dataset: its kinematic coverage, measured central values, and associated uncertainties.
The sensitivity of QCFs, model parameters, or other derived quantities to these data properties may be probed by perturbing the data and observing corresponding shifts in the model quantities.

In the Hessian formalism, our model is specified by identifying the parameter vector corresponding to the maximum likelihood estimator (MLE), denoted $\hat{\vb*{a}}(\mathcal{D})$ and defined by the condition $\grad_{\vb*{a}} \chi^2(\hat{\vb*{a}}(\mathcal{D}),\mathcal{D}) = 0$.
The uncertainties of our model are provided by expanding the $\chi^2$ to quadratic order, rendering the likelihood as a multivariate Gaussian with parameter covariance matrix $\Sigma \!=\! {\rm H}^{-1}$, where ${\rm H}_{ij} \!=\! \tfrac{1}{2} \partial^2 \chi^2(\vb*{a},\mathcal{D}) / \partial a_i \, \partial a_j |_{\vb*{a} = \hat{\vb*{a}}(\mathcal{D})}$ is the Hessian matrix evaluated at the MLE.
Although $\hat{\vb*{a}}$ and $\Sigma$ are generally nonlinear functions of the data, infinitesimal perturbations to the dataset $\delta \mathcal{D}$ induce linear responses in extracted model quantities, where the coefficients of these small shifts define model sensitivities to data~\cite{simonyan2014deepinsideconvolutionalnetworks,giordano2015linear}.

Let $\lambda$ denote an arbitrary property of the data.
Differentiating the condition defining the MLE with respect to $\lambda$, we obtain an equation which gives 
\begin{align}
\label{eq:mle-response}
    \frac{\dd \hat{a}_{i}}{\dd \lambda} = - \frac{1}{2} \sum_{j} {\rm H}^{-1}_{ij} \frac{\partial^{2} \chi^2}{\partial a_{j} \, \partial \lambda}
,\end{align}
implying that the sensitivity of a model parameter to data perturbations is determined through the combination of the local curvature of the $\chi^2$ and the joint variation of the $\chi^2$ with respect to parameters and the data property.
As a concrete example, take $\lambda \!=\! D_k$ at fixed $V$, which gives $\partial^2 \chi^2 / \partial a_j \partial D_k = -2 \sum_l (V^{-1})_{kl} (\partial T_l / \partial a_j)$ and reduces to $- \tfrac{2}{\sigma_k^2} (\partial T_k/\partial a_j)$ for diagonal uncertainties.
The response of any derived quantity $F(\vb*{a})$ follows immediately from the chain rule
\begin{align}
\label{eq:F-MLE-response}
    \frac{\dd F(\hat{\vb*{a}})}{\dd \lambda} = \frac{\partial F(\hat{\vb*{a}})}{\partial \lambda} - \frac{1}{2} \sum_{ij} \frac{\partial F}{\partial a_i} {\rm H}^{-1}_{ij} \frac{\partial^2 \chi^2}{\partial a_j \, \partial \lambda} 
,\end{align}
which couples the variation of $F$ from parameter and data shifts to the local curvature of the $\chi^2$.
This result provides a direct mapping from infinitesimal perturbations of experimental information to shifts in fitted parameters and any derived quantities, making explicit the flow of information from data space to model space.
Note that the dependence in Eq.~(\ref{eq:F-MLE-response}) on $\lambda$ is left general since quantities such as the full $\chi^2$ depend explicitly on data properties.
However, the explicit partial derivative term in Eq.~(\ref{eq:F-MLE-response}) vanishes in practice for quantities such as QCFs which inherit $\lambda$ dependence only implicitly through the MLE.

Beyond central values, the parameter covariance matrix in the local Hessian approximation is defined by the condition $\Sigma {\rm H} \!=\! \mathds{1}$.
Differentiating this relation with respect to $\lambda$ yields
\begin{align}
\label{eq:hessian-covmat-response}
    \frac{\dd \Sigma}{\dd \lambda} = - \Sigma \frac{\dd {\rm H}}{\dd \lambda} \Sigma
.\end{align}
Gradients of the parameter uncertainties and correlations are then given, respectively, by
\begin{align}
\label{eq:hessian-unc-corr-response}
\begin{aligned}
    \frac{\dd{(\delta a_{i})}}{\dd{\lambda}} &= \frac{1}{2 \delta a_i} \frac{\dd{\Sigma_{ii}}}{\dd{\lambda}} \\
    \frac{\dd{\rho_{ij}}}{\dd{\lambda}} &= \rho_{ij} \, \Bigg( \frac{1}{\Sigma_{ij}} \frac{\dd{\Sigma_{ij}}}{\dd{\lambda}} - \frac{1}{2 \Sigma_{ii}} \frac{\dd{\Sigma_{ii}}}{\dd{\lambda}} - \frac{1}{2 \Sigma_{jj}} \frac{\dd{\Sigma_{jj}}}{\dd{\lambda}} \Bigg) \,
,\end{aligned}
\end{align}
where $\rho_{ij} \!=\! \Sigma_{ij} / \sqrt{ \Sigma_{ii} \Sigma_{jj}}$ is the correlation between parameters $a_i$ and $a_j$.
Eqs.~(\ref{eq:hessian-covmat-response}) and (\ref{eq:hessian-unc-corr-response}) demonstrate that the linear response formalism provides a consistent treatment of data sensitivities for model uncertainties and correlations in addition to central values.
This may be particularly useful for phenomenological impact studies of future experiments, where the primary goal is to understand how such data may constrain QCFs~\cite{Whitehill:2026swa,Sawaya:2026dqz}.

The response functions above can also be formulated for Bayesian global analyses, where the goal is to characterize the posterior distribution of model parameters given data.
In this approach, the primary object is not a set of estimators but a data-dependent probability measure on parameter space.
Perturbations of the data therefore modify the posterior distribution itself, redistributing probability mass across parameter space.
Aggregate sensitivities to this redistribution can be accessed through gradients of statistical summaries, such as posterior means and covariances, with respect to properties of the data.

We denote a general posterior distribution $p(\vb*{a}|\mathcal{D}) = \mathcal{L}(\vb*{a},\mathcal{D}) \pi(\vb*{a}) / Z(\mathcal{D})$, where $\pi(\vb*{a})$ is a prior distribution of parameters (assumed to be independent of the data) and $Z(\mathcal{D})$ serves as a normalization.
The derivative of the posterior with respect to a data property $\lambda$
\begin{align}
\label{eq:posterior-response}
    \frac{\dd p(\vb*{a} | \mathcal{D})}{\dd \lambda} = - \frac{1}{2} \Bigg[ \frac{\dd \chi^2}{\dd \lambda} - \Big\langle \frac{\dd \chi^2}{\dd \lambda} \Big\rangle \Bigg] p(\vb*{a}|\mathcal{D})
,\end{align}
where we note that $\vb*{a}$ is an integration variable here, independent of $\lambda$, so $\dd{\chi^2} \!/\! \dd{\lambda}$ contains only the explicit dependence of $\chi^2$ on the data.
From this, we can see the mean of any derived quantity $F$ has derivative
\begin{align}
\label{eq:F-bayesian-response}
    \frac{\dd \expval{F}}{\dd \lambda} = \Big\langle \frac{\partial F}{\partial \lambda} \Big\rangle - \frac{1}{2} {\rm Cov}\Big( F, \frac{\dd \chi^2}{\dd \lambda} \Big)
,\end{align}
showing that sensitivity is governed by the covariance between model quantities and data-induced variations of $\chi^2$.
This expression is an application of linear response theory and statistical mechanics to Bayesian inference~\cite{Ruelle_2009,baehrens2009explainindividualclassificationdecisions,giordano2015linear}: the likelihood takes the form of a Gibbs distribution, where $\chi^2/2$ serves as the data-dependent contribution to the energy associated with a given parameter configuration.
In this picture, we can view Eq.~(\ref{eq:F-bayesian-response}) as defining a susceptibility of a model quantity to data perturbations, which act like external drivers of the posterior distribution.

The response of the mean parameters is obtained directly by replacing $F(\vb*{a}) \!=\! \vb*{a}$, and the response of ${\rm Cov}(a_i,a_j)$ follows with the replacement $F(\vb*{a}) \!=\! (a_i - \langle a_i \rangle) (a_j - \langle a_j \rangle)$.
After linearizing $F$ in parameter space under the Hessian approximation~\cite{Pumplin:2001ct}, Eq. (\ref{eq:F-bayesian-response}) reproduces the Hessian response functions in Eqs. (\ref{eq:mle-response})--(\ref{eq:hessian-covmat-response}), establishing a consistent linear-response structure for both Hessian and Bayesian analyses.

\subsection{Influence functions} 
\label{ssec:influence-functions}

Another matter of interest is understanding how much a model's behavior is influenced by the presence of a data point or dataset.
In principle, one can remove data points or datasets from $\mathcal{D}$ and run new analyses to assess their influence, but in practice, the complexity and scale of modern global analyses make this proposition an expensive and time-consuming endeavor.
Using the technology of influence functions, one can efficiently approximate the results of this procedure using the results of a single analysis~\cite{koh2017understanding,lee2025influencedynamicsstagewisedata,giordano2019swiss}.

We consider decreasing the weight of a data point $D_j$ in the likelihood by a small amount $\epsilon$ so that $(\chi^2)' = \chi^2 - \epsilon \chi^2_{j}$, where $\chi^2_j$ is the contribution from the $j^{\rm th}$ data point to $\chi^2$ and removing the data point amounts to setting $\epsilon \!=\! 1$.
The results for the response functions carry over directly to the influence functions $\mathcal{I}_{\langle F \rangle,j} = \dd{\langle F \rangle} / \dd{\epsilon}$ with the replacement $\lambda \!\rightarrow\! \epsilon$, in which case we have $\dd{(\chi^2)'} / \dd{\epsilon} = -\chi^2_j$.
Thus, within the local quadratic Hessian approximation for the likelihood, the influence functions for the Hessian MLE and covariance
\begin{align}
\label{eq:hessian-influences}
\begin{aligned}
    \mathcal{I}_{F(\hat{\vb*{a}}),j} &= \frac{1}{2} \sum_{lm} \frac{\partial F}{\partial a_l} {\rm H}^{-1}_{lm} \pdv{\chi^2_{j}}{a_{m}} \\
    \mathcal{I}_{\Sigma(\hat{\vb*{a}}),j} &= \Sigma {\rm H}^{(j)} \Sigma
,\end{aligned}
\end{align}
respectively, where ${\rm H}^{(j)}$ is the Hessian of $\chi^2_{j}$, defined with the same normalization as ${\rm H}$ and evaluated at the MLE.
The Bayesian influence functions follow similarly, with Eq. (\ref{eq:F-bayesian-response}) becoming
\begin{align}
\label{eq:F-bayesian-influence}
    \mathcal{I}_{\langle F \rangle,j} = \frac{1}{2} {\rm Cov}(F,\chi^2_j)
.\end{align}
Note that the partial derivative term from Eq.~(\ref{eq:F-bayesian-response}) vanishes since $\epsilon$ appears only at the level of the reweighted $(\chi^2)'$ and not directly in any model quantities obtained using $\chi^2$.

We observe that the influence function for $\langle F \rangle$ is related to the $L_2$ sensitivity~\cite{Hobbs:2019gob,Jing:2023isu} as
\begin{align}
    \mathcal{I}_{\langle F \rangle,j} = \frac{\delta F}{2} S_{F,L_2}(j)
.\end{align}
Originally, the $L_2$ sensitivity was defined in the context of Hessian analyses as the directional derivative of an experimental $\chi^2$ along the unit direction where $F$ varies most strongly.
It also admits a Bayesian interpretation.
If the prior $\pi(\vb*{a})$ includes a factor $\exp\big( \eta F(\vb*{a}) \big)$, then $S_{F,L_{2}}(j) \!=\! \tfrac{1}{\delta F} \dd{\langle \chi^2_{j} \rangle} / \dd{\eta}|_{\eta=0}$.
Thus, while the $L_2$ sensitivity and influence function probe the same local covariation of $F$ and $\chi^2_j$, their distinction lies in which quantity is perturbed and which response is measured.
The $L_2$ sensitivity quantifies how $\chi^2_j$ responds to a displacement toward larger $F$, whereas the influence function characterizes how $F$ responds to a perturbation of the statistical weight of data point $j$.

Defining $\chi^2_j$ in practice is straightforward when data are uncorrelated.
In such a case, each data point defines an independent statistical mode so that removing data point $j$ is equivalent to subtracting its squared residual from the total $\chi^2$.
Data correlations generally break this equivalence, rendering the definition of $\chi^2_j$ non-unique.
Establishing a precise definition for $\chi^2_j$ in their presence then depends on the particular attribution being sought, requiring one to specify which aspect of the likelihood is being perturbed.

A strict leave-one-out (LOO) attribution may be defined through $\chi^2_j \equiv \chi^2 - \chi^2_{-j}$, where $\chi^2_{-j}$ is obtained by removing data point $j$ from its dataset.
For Gaussian data, this difference represents the conditional contribution of point $j$ to the likelihood given the remaining measurements.
Such contributions, however, do not form an additive decomposition of the total $\chi^2$.
If such a decomposition is desired, the resulting terms generally mix information from correlated measurements and need not retain the direct physical interpretation of the original data points.
In a nuisance-parameter representation, as in Ref.~\cite{Stump:2001gu}, the profiled likelihood may be written as a sum of squared shifted residuals together with penalty terms for the nuisance parameters.
The shifted residuals provide additive pointwise contributions, although downweighting their contribution corresponds to perturbing effective shifted data points rather than the original measurements.
Alternatively, in a covariance-matrix treatment, one may choose a whitening matrix $W$ satisfying $V^{-1} \!=\! W^{\rm T} W$ so that $\chi^2 = || W \vb*{\Delta} ||^{2}$ is a sum of squared residuals, although $W$ is not necessarily unique and generally defines transformed statistical modes rather than individual physical data points.
These ambiguities can be avoided at the level of independent experimental datasets by using the total experimental $\chi^2_E$ in place of $\chi^2_j$ in Eqs. (\ref{eq:hessian-influences}) and (\ref{eq:F-bayesian-influence}).
This approach mirrors previous studies using the $L_2$ sensitivity~\cite{Wang:2018heo,Jing:2023isu}, though one then probes the influence of an entire dataset rather than its individual points.

\subsection{Interpretations and practical considerations}
\label{ssec:interpretations-and-practical-considerations}

Response and influence functions can generally be computed as post-fit diagnostics.
In Hessian analyses, the required parameter derivatives may be approximated using finite differences along the eigenvector directions, and in Bayesian analyses they may be obtained from sample covariances over the posterior ensemble.
This permits their calculation for many data points and observables without repeating the full optimization or sampling procedure or performing a separate reweighting for each perturbation, offering relatively inexpensive computational means to probe data sensitivities.

While the response and influence functions defined above are exact local differential quantities, their practical interpretation is naturally tied to model responses under finite perturbations of the data:
\begin{align}
\label{eq:response-influence-interpretation}
\begin{aligned}
    \langle F \rangle(\lambda + \delta \lambda) &= \langle F \rangle(\lambda) + \delta \lambda \frac{\dd{\langle F \rangle}}{\dd{\lambda}} + \mathcal{O}(\delta \lambda^2) \\
    \langle F \rangle(\epsilon) &= \langle F \rangle(0) + \epsilon \, \mathcal{I}_{\langle F \rangle,j} + \mathcal{O}(\epsilon^2)
,\end{aligned}
\end{align}
respectively, for response and influence functions.
Thus, these functions define the coefficients of the local linear terms.
Their use to predict finite responses under finite shifts $\delta \lambda$ or extrapolating local data influences to complete data removal at $\epsilon \!=\! 1$ introduces an additional approximation from neglecting higher-order terms, and its range of validity depends on the fitted quantity and perturbation considered.
This may be tested by comparing reweighted or refit results at varying perturbation sizes with a linear trend, where deviations signal higher-order effects.

In Bayesian applications, response and influence functions are estimated from sample covariances and therefore carry finite-sample uncertainty.
This uncertainty generally scales as $1/\sqrt{N}$, where $N$ is the effective number of independent samples, and may be computed analytically in simple cases or estimated through bootstrap methods, for example, for a general posterior.
Separately, when a response or influence is interpreted as a finite shift of a posterior summary $\langle F \rangle$, its magnitude must be compared with the sampling uncertainty of the estimator for $\langle F \rangle$.
Predicted shifts below this scale cannot be reliably distinguished from ordinary finite-sample variation.

\section{Toy DIS application} 
\label{sec:toy-dis-application}

\begin{figure}[t]
    \centering
    \includegraphics[width=\linewidth]{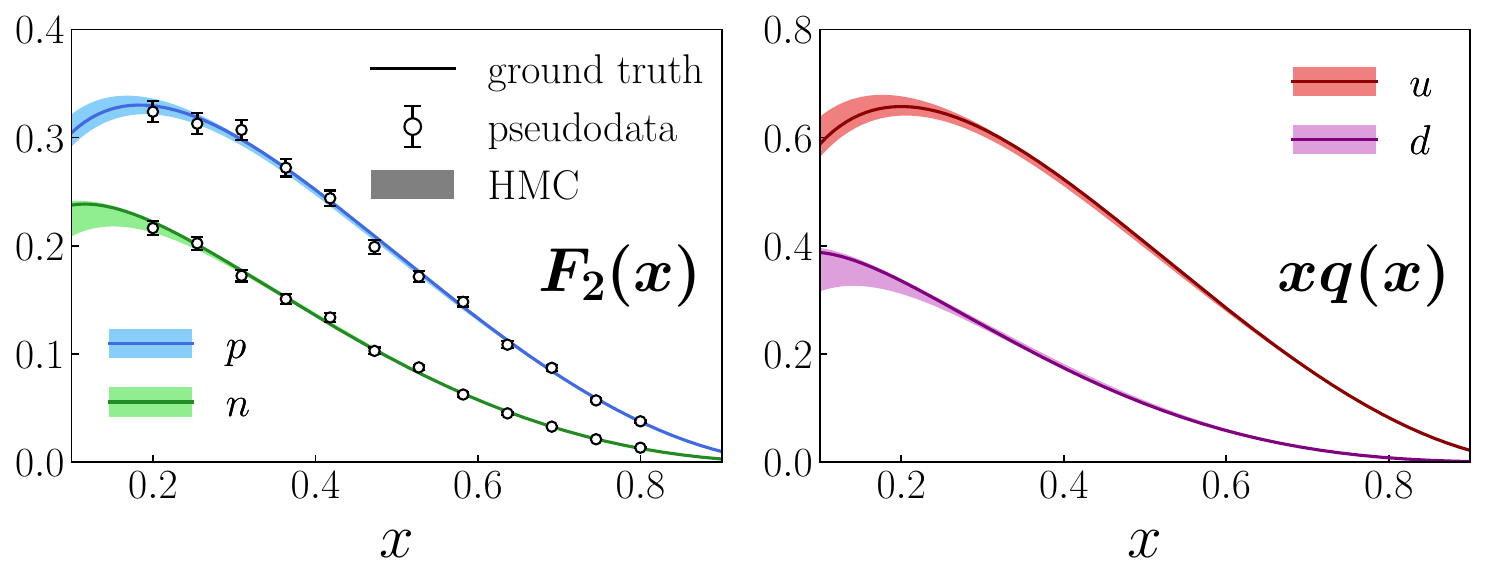}
    \caption{Hamiltonian Monte Carlo (HMC) results for the baseline pseudodata analysis. \textbf{Left}: Proton ($p$) and neutron ($n$) $F_2$ pseudodata compared to the HMC posterior mean with a $1\sigma$ envelope and the ground truth. \textbf{Right}: Up ($u$) and down ($d$) PDFs $x q(x)$ from the same posterior, showing the extracted mean and $1\sigma$ envelope alongside the ground truth.}
    \label{fig:fit}
\end{figure}

\begin{figure*}[t]
    \centering
    \includegraphics[width=0.75\linewidth]{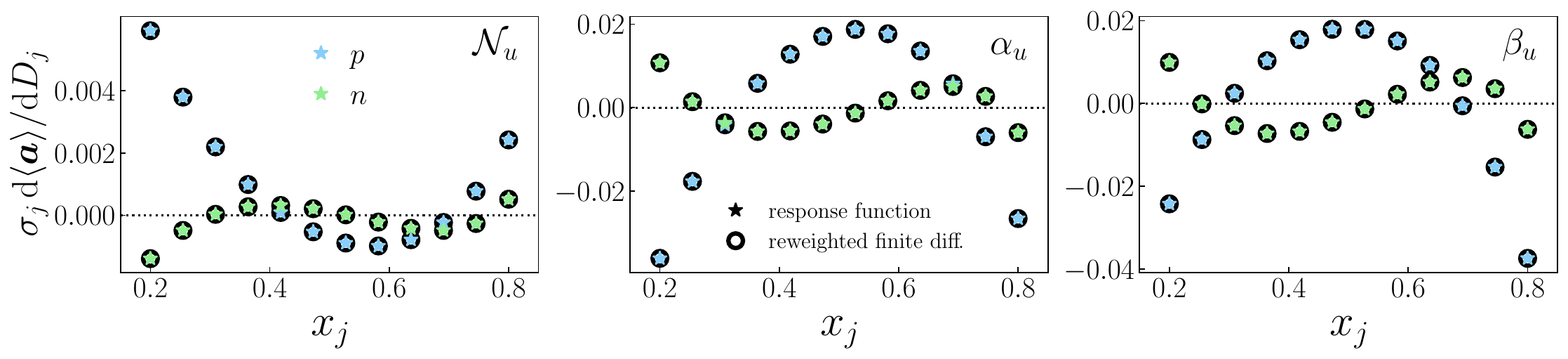}
    \par
    \includegraphics[width=0.5\linewidth]{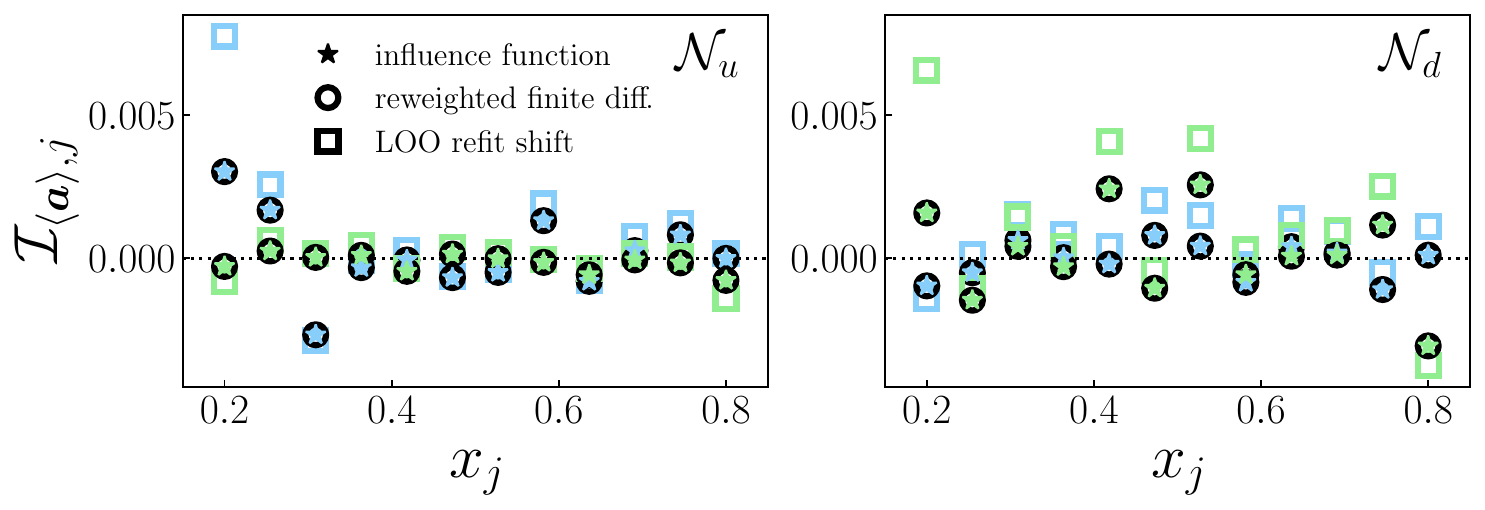}
    \caption{Validation of parameter-level response and influence functions. \textbf{Top}: Scaled response functions $\sigma_j \dd{\langle \vb*{a} \rangle} / \dd{D_j}$ for $\mathcal{N}_u$, $\alpha_u$, and $\beta_u$, computed using Eq.~(\ref{eq:F-bayesian-response}) and compared with reweighted finite differences for proton ($p$) and neutron ($n$) pseudodata. \textbf{Bottom}: Influence functions $\mathcal{I}_{\langle \vb*{a} \rangle,j}$ for up and down quark normalization parameters compared with reweighted finite differences and explicit leave-one-out refit (LOO) shifts.}
    \label{fig:a-response-influence}
\end{figure*}

As a simplified concrete application of the response and influence functions, we perform an extraction of the up and down PDFs from inclusive DIS pseudodata on the parton model $F_2$ structure function~\cite{Feynman:1969ej,Bjorken:1968dy,Bjorken:1969ja,Callan:1968zza,Blumlein:2012bf}.
For flavor separation, we use both proton and neutron pseudodata, given by
\begin{align}
\label{eq:F2-observable}
\begin{aligned}
    F_2^p(x) &= \tfrac{4}{9} xu(x) + \tfrac{1}{9}xd(x) \\
    F_2^n(x) &= \tfrac{4}{9} xd(x) + \tfrac{1}{9} xu(x)
,\end{aligned}
\end{align}
where the latter assumes isospin symmetry.
The PDFs, neglecting evolution, are modeled using a standard parametric form
\begin{align}
\label{eq:pdf-model}
    q(x) = \frac{\mathcal{N}_{q} x^{\alpha_q} (1 - x)^{\beta_q}}{{\rm B}(\alpha_q + 2, \beta_q + 1)}
.\end{align}

We generate pseudodata from a set of ``ground-truth'' parameters across 12 uniformly spaced points between $x \!=\! 0.2$ and $x \!=\! 0.8$ for each target.
Each point is assigned a $3\%$ uncorrelated uncertainty relative to the true value, and the pseudodata central values are defined by resampling the true value within uncertainties.
Our $\chi^2$ is defined as
\begin{align}
\label{eq:chi2-toy-def}
    \chi^2 = \sum_{e} \sum_{i} \Big( \frac{D_{e,i} - F_2(x_{e,i},\vb*{a})}{\sigma_{e,i}} \Big)^2
,\end{align}
where $e$ denotes the target index, $i$ denotes the data point index, and $\sigma_{e,i}$ denotes the corresponding uncorrelated uncertainty.
Using a Hamiltonian Monte Carlo (HMC) algorithm~\cite{cobb2021scaling}, we draw 10,000 independent samples, which generally limits finite-sample noise in our demonstration below, from the posterior under a weak uniform prior.
Results for $F_2$ and the PDFs are shown in Fig.~\ref{fig:fit}.
The pseudodata are well described by our model, and the ground truth PDFs and $F_2$ observables are recovered within the $1\sigma$ error bands, establishing a controlled baseline that permits direct interpretation of the response and influence functions below.

Before examining data sensitivities at the level of the PDFs themselves, we first validate the response and influence functions at the level of the fitted parameters.
In Fig.~\ref{fig:a-response-influence}, we verify that the Bayesian covariance calculation of the gradients in Eqs.~(\ref{eq:F-bayesian-response}) and~(\ref{eq:F-bayesian-influence}) matches those from reweighted finite differences from perturbing the data central values and likelihood weights for the scaled response and influence functions, respectively.

\begin{figure}[t]
    \centering
    \includegraphics[width=\linewidth]{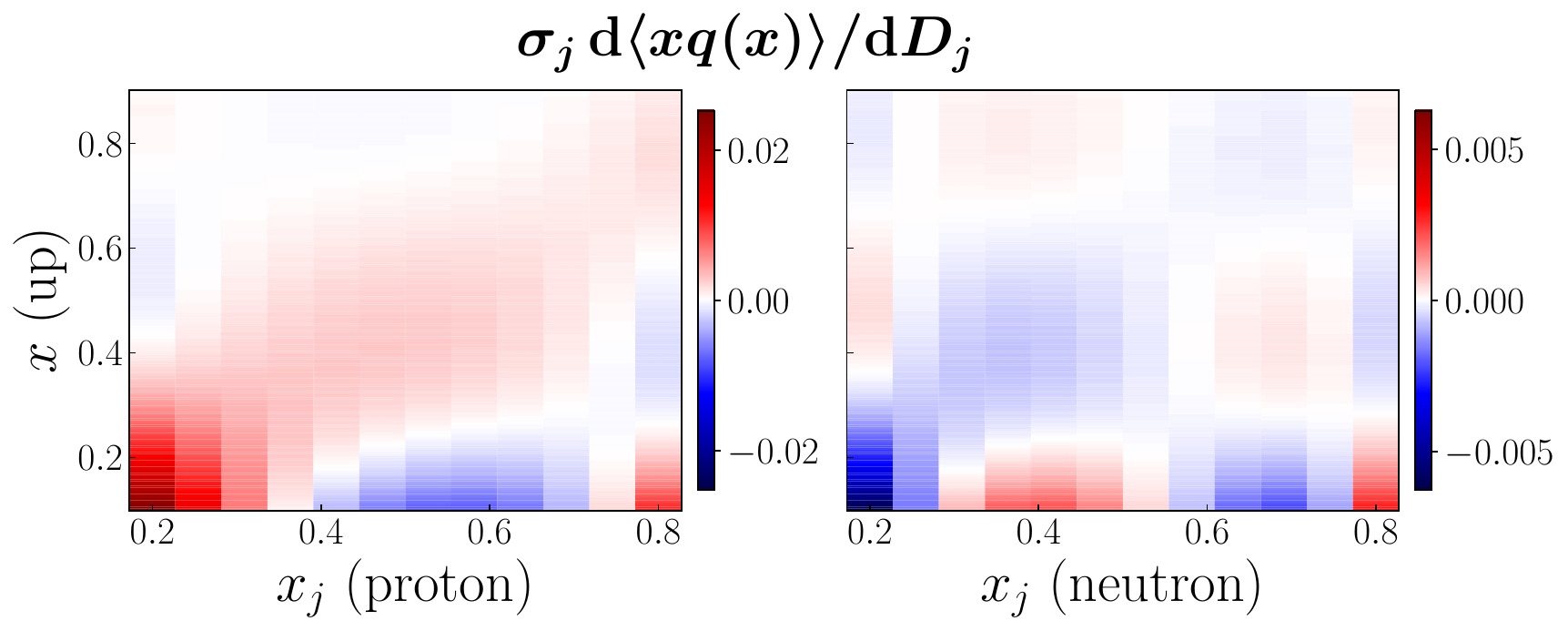}
    \caption{Response of up-quark PDF to proton and neutron pseudodata, scaled by data uncertainties. Each panel shows $\sigma_j \, \dd \langle x u(x)\rangle / \dd D_j$ as a function of the PDF momentum fraction $x$ and pseudodata kinematic position $x_j$.}
    \label{fig:xq-y-grad}
\end{figure}

\begin{figure}[t]
    \centering
    \includegraphics[width=\linewidth]{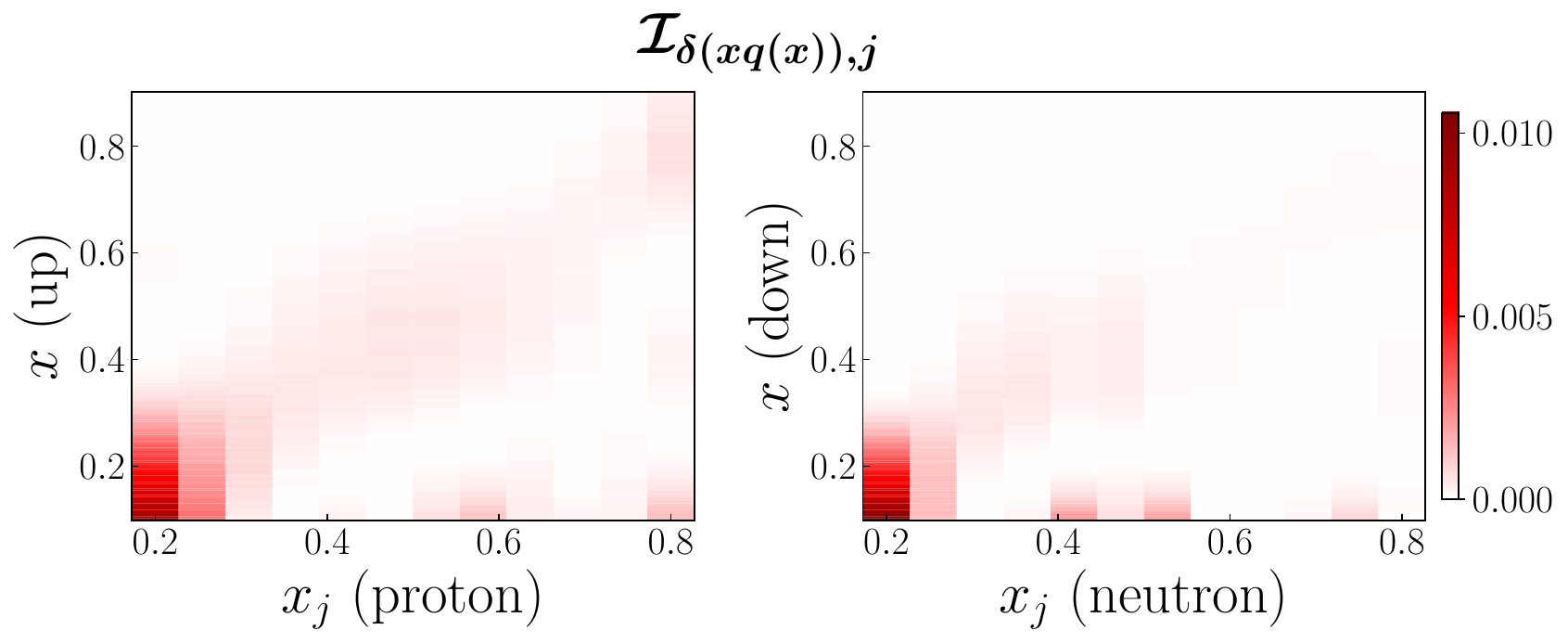}
    \caption{Influence of proton and neutron pseudodata on the up and down-quark PDF uncertainties (left and right, respectively).
    }
    \label{fig:xq-dxq-influence}
\end{figure}

\begin{figure*}[t]
    \centering
    \includegraphics[width=0.9\linewidth]{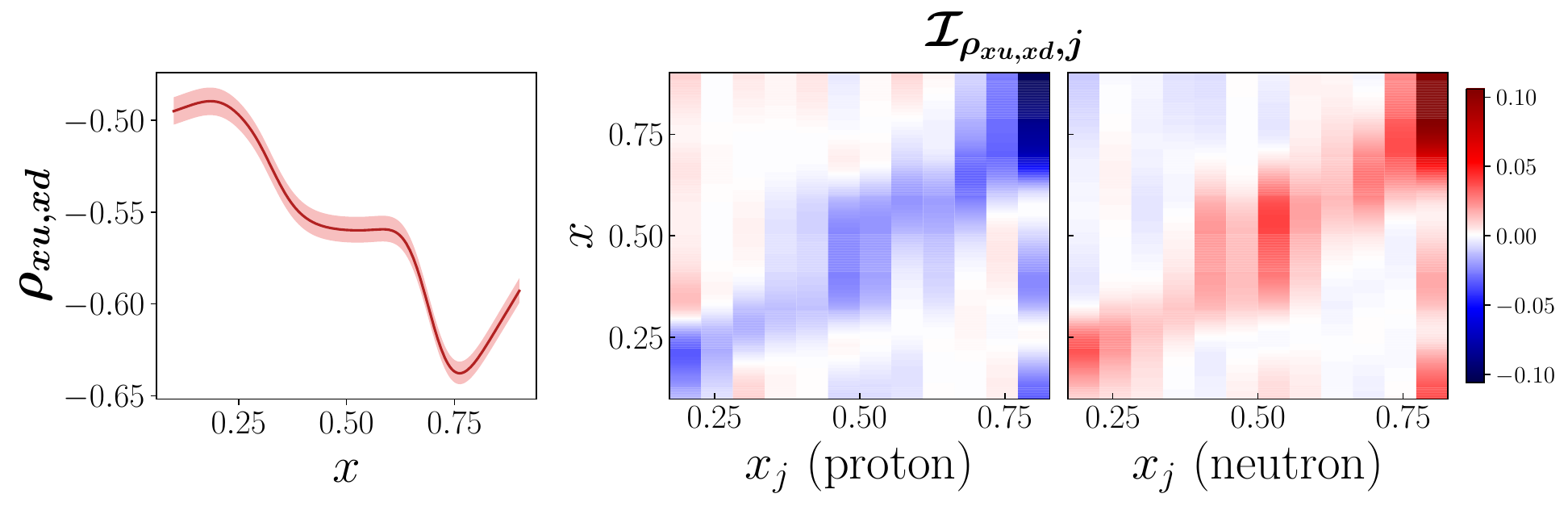}
    \caption{Influence of pseudodata on the up–down PDF correlation $\rho_{xu,xd}$. \textbf{Left}: Correlation $\rho_{xu,xd}(x)$ as a function of $x$, where the band shows the corresponding standard error, estimated via bootstrap. \textbf{Right}: Influence function $\mathcal{I}_{\rho_{xu,xd},j}$ for proton and neutron pseudodata. The sign indicates whether downweighting a data point locally strengthens or weakens the up-down correlation.}
    \label{fig:corrud-influence}
\end{figure*}

Furthermore, alongside the influence, we show the finite shifts in the parameter means obtained from a LOO refit, in which a pseudo-datapoint $j$ is removed by restricting the sum in Eq.~(\ref{eq:chi2-toy-def}) to $i \!\ne\! j$ for a particular target.
In terms of Eq.~(\ref{eq:response-influence-interpretation}), the LOO refit shift is the difference between $\langle F \rangle$ at $\epsilon \!=\! 1$ and the original, unperturbed result with $\epsilon \!=\! 0$.
These explicit LOO refits test the first-order extrapolation from an infinitesimal reweighting to complete removal of a data point.
Strong deviations from the linear influence-function prediction indicate the importance of missing higher-order dependence on $\epsilon$, as is generally evident for the down-quark normalization parameter and in a few instances for the up-quark normalization parameter.

The results also reproduce physically expected trends.
The displayed up-quark responses show larger sensitivity to proton measurements.
The shape parameters also exhibit the expected kinematic sensitivity to the lower- and higher-$x_j$ regions, respectively, while retaining appreciable sensitivity to intermediate-$x_j$ pseudodata.
The latter reflects in part how variations of the shape parameters redistribute the PDF across $x$ through the beta-function normalization while preserving $\mathcal{N}_q$ as an integrated moment.

Having validated the response and influence functions at the parameter level, we now turn to the PDFs, which are more directly the quantities of physical interest.
Fig.~\ref{fig:xq-y-grad} maps the scaled response of the up-quark PDF to perturbations in the proton and neutron pseudodata central values.
The dominant response lies along the diagonal $x \!\approx\! x_j$.
This is largely a manifestation of parton model features: the charge-weighted linear combinations of the PDFs in $F_2$ and the identification of the Bjorken scaling variable and the parton momentum fraction.
Large, positive responses to proton pseudodata and small, negative responses to neutron pseudodata for the up quark PDF are observed, and we also see that the scaled response function is larger at smaller $x$ and $x_j$, which can be partly attributed to the relatively larger size of the PDFs and larger absolute uncertainties of the pseudodata in those regions, respectively.

In modern global analyses, understanding how data shapes the posterior geometry of extracted QCFs is an essential complement to understanding how data influences the central values of QCFs.
Fig.~\ref{fig:xq-dxq-influence} shows the influence of proton and neutron pseudodata on the up and down quark PDF uncertainties, respectively.
In fact, for the diagonal covariance used here, the influence function $\mathcal{I}_{\delta(x q(x)),j}$ is equivalent to the scaled response function $\sigma_j^2 \dd{[\delta(x q(x))]} / \dd{\sigma_j^2}$.
The maps for both flavor and target combinations display primarily positive values, with a small subset of negative values (not shown in the plot) lying below the estimated noise threshold.
This indicates that losing information from the pseudodata generally leads to increased uncertainties in the extracted PDFs, as expected for our low-dimensional fit to mutually consistent pseudodata.
Furthermore, we observe that the influence of the pseudodata on the PDF uncertainties generally decreases away from the diagonal $x \!\approx\! x_j$, arising primarily from our use of the parton model.

Beyond the marginal PDF uncertainties considered above, the same framework can probe how data constrain joint posterior structure.
We illustrate this in Fig.~\ref{fig:corrud-influence} by examining the influence of $F_2$ pseudodata on the correlation between the up and down PDFs.
The left panel shows the strong negative correlation between the up and down PDFs, together with its bootstrap-estimated standard error, as a function of $x$ and the corresponding influences to proton and neutron pseudodata are shown in the two right panels.
The proton influence displays a strong negative diagonal while the neutron influence appears flipped in sign but roughly equal in magnitude.
We can intuitively understand this as follows.
Pseudodata on $F_2^p$ is dominated by the value of $u$, permitting a relatively independent determination of the up-quark distribution. 
On the other hand, the $u$ and $d$ contributions to $F_2^n$ are more balanced, coupling the flavors more strongly.
Therefore, downweighting or removing proton pseudodata strengthens the anti-correlation between the up and down PDFs, whereas removing neutron pseudodata weakens it.

\begin{figure}[t]
    \centering
    \includegraphics[width=\linewidth]{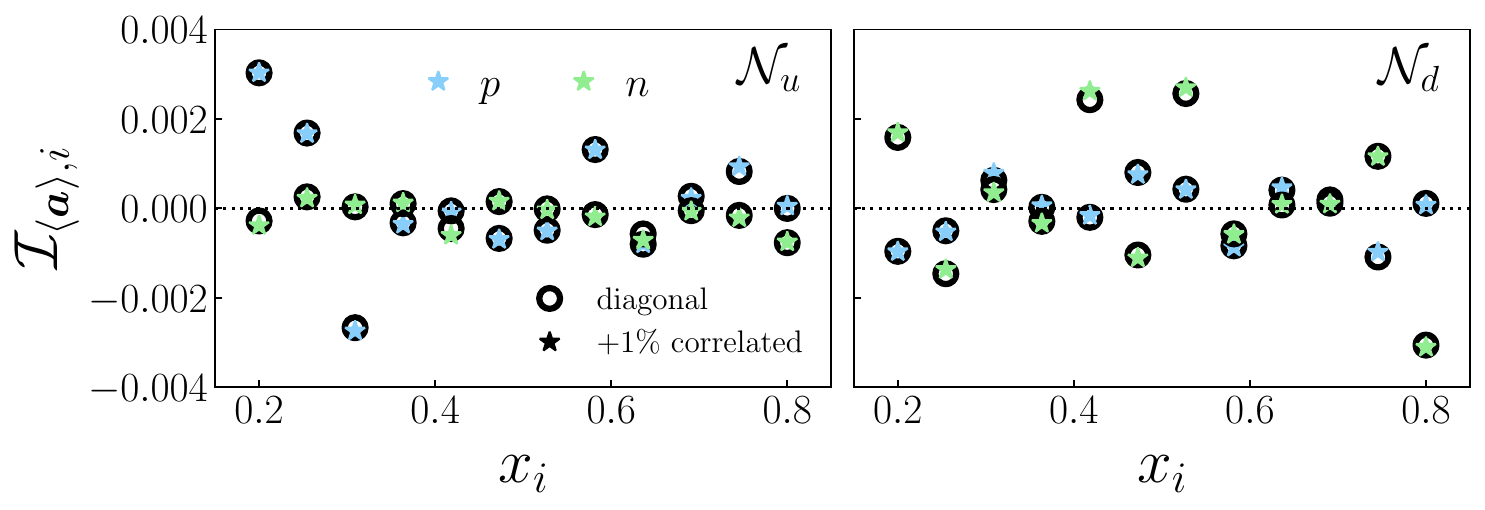}
    \caption{Influence of proton and neutron pseudodata on the up- and down-quark normalization parameters, $\mathcal N_u$ and \(\mathcal N_d\). Open circles show the baseline diagonal-covariance results, while stars include an additional $1\%$ fully correlated uncertainty relative to the ground truth for each target dataset, with $\chi_j^2 \equiv \chi^2-\chi_{-j}^2$.}
    \label{fig:influence-w-corr}
\end{figure}

Finally, we provide an explicit example of the influence function in the presence of correlated data uncertainties.
Specifically, we supplement the 3\% uncorrelated uncertainty used in the baseline analysis with an additional 1\% fully correlated uncertainty, relative to the ground truth, for each target pseudo-dataset.
Fig.~\ref{fig:influence-w-corr} compares the influence of proton and neutron pseudodata on the up- and down-quark normalization parameters for the diagonal and correlated covariance treatments.
In particular, for the latter case, we use the LOO definition for $\chi^2_j$, although we have verified that alternative definitions give qualitatively similar results.
The resulting influence functions are modified only modestly, as expected for the relatively small correlated contribution considered here, and retain the same overall sensitivity patterns.
This example illustrates that local influence functions can be evaluated for correlated likelihoods once the perturbation and its attribution within the likelihood are specified.

\section{Conclusion}
\label{sec:conclusion}

In this work, we adapt the use of linear response and influence functions for characterizing data sensitivity in Hessian and Bayesian global QCD analyses and demonstrate their application through a controlled PDF extraction from DIS pseudodata.
These quantities describe the response of QCF central values, uncertainties, and their correlations under perturbations of experimental inputs or their statistical weight, thereby providing differential maps between experimental measurements and the statistical properties of extracted QCFs.

Scaling this framework to realistic global analyses will introduce additional theoretical and model complexity, including QCD evolution, convolutional relations between QCFs and observables, and a broader set of QCFs involving additional partonic flavors and more flexible parameterizations.
While such features may complicate the evaluation and interpretation of the resulting sensitivities, they should not alter the basic response and influence framework presented here.
Several related questions and extensions nevertheless require further study.
In particular, the definitions and interpretations of point-level influence in the presence of correlated data should be explored more systematically, including the practical consequences of different choices for likelihood attribution.
The framework may also prove useful for diagnosing data tensions, informing dataset selection, and identifying kinematic regions where future measurements may have the greatest impact.
A further direction is the formulation of analogous sensitivity metrics for global analyses based on data resampling or replica methods, where the relevant sensitivities would characterize distributions of estimators rather than a posterior distribution.
Detailed studies of these applications and extensions are left for future work.

Faithfully representing and interpreting the connection between experimental data and QCFs in the presence of complex phenomenological and theoretical models and large datasets remains an important challenge for global QCD analysis.
Linear response and influence functions broaden the available toolkit for addressing this challenge, complementing existing approaches to quantifying experimental sensitivity and impact.
The combined use and further development of such diagnostics should become increasingly valuable as theoretical precision improves and as measurements at Jefferson Lab, the Electron-Ion Collider, and the Large Hadron Collider further expand the kinematic reach and statistical precision of experimental data.

\section*{Acknowledgements}

The author thanks Arkaitz Rodas-Bilbao for helpful discussions and comments during the preparation of this manuscript.
The work of R.~M.~W. was partially supported by the Jefferson Science Associates (JSA) Graduate Fellowship.
This material is based upon work supported by the U.S. Department of Energy, Office of Science, Office of Workforce Development for Teachers and Scientists, Office of Science Graduate Student Research (SCGSR) program. The SCGSR program is administered by the Oak Ridge Institute for Science and Education (ORISE) for the DOE. 
ORISE is managed by ORAU under contract number
DESC0014664. 
All opinions expressed in this paper are the author’s and do not necessarily reflect the policies and views of DOE, ORAU, or ORISE.
This material is based upon work supported by the U.S. Department of Energy, Office of Science, Office of Nuclear Physics under Contract No. 89243126CSC000213.

\bibliography{bibliography}

\end{document}